\documentclass[a4paper,11pt]{article}
\usepackage{aaskaiid}
\usepackage{import}
\usepackage{hyperref}
\usepackage{orcidlink}

\def\gtrsim{\mathrel{\hbox{\rlap{\hbox{%
 \lower4pt\hbox{$\sim$}}}\hbox{$>$}}}}

\title{SKA--VLBI view of AGN jets in the early Universe}
\ShortTitle{SKA--VLBI view of AGN jets in the early Universe}

\author[1]{\textbf{C.~Spingola\orcidlink{0000-0002-2231-6861}}}
\ShortName{C.~Spingola et al.} % shortened name list for header 
\author[2,3]{\textbf{M.~Mezcua\orcidlink{0000-0003-4440-259X}}}
\author[4]{\textbf{Y.~Liu}}
\author[5,6]{\textbf{S.~Frey}}
\author[7,8]{\textbf{S.~Belladitta\orcidlink{0000-0003-4747-4484}}}
\author[23,4]{\textbf{T.~An\orcidlink{0000-0003-4341-0029}}}
\author[1]{M.~Giroletti\orcidlink{0000-0002-8657-8852}}
\author[9]{A.~Caccianiga\orcidlink{0000-0002-2339-8264}} 
\author[9]{A.~Moretti\orcidlink{0000-0002-9770-0315}} 
\author[9,10]{L.~Ighina\orcidlink{0000-0003-1516-9450}}
\author[9]{T.~Sbarrato\orcidlink{0000-0002-3069-9399}} 
\author[1]{G.~Migliori\orcidlink{0000-0003-0216-8053}}
\author[1,11]{D.~Dallacasa\orcidlink{0000-0003-1246-6492}}
\author[12,13]{L.~I.~Gurvits\orcidlink{0000-0002-0694-2459}} 
\author[5,14,15,16]{K.~\'E.~Gab\'anyi\orcidlink{0000-0003-1020-1597}}
\author[5]{K.~Perger\orcidlink{0000-0002-6044-6069}} 
\author[12]{Z.~Paragi\orcidlink{0000-0002-5195-335X}} 
\author[1]{R.~D.~Baldi\orcidlink{0000-0002-1824-0411}}
\author[7]{E.~Ba\~{n}ados\orcidlink{0000-0002-2931-7824}}
\author[1]{G.~Bernardi\orcidlink{0000-0002-0916-7443}}  
\author[1]{M.~Bondi}
\author[1,11]{E.~De~Rubeis\orcidlink{0000-0002-0428-2055}}  
\author[1]{F.~D'Ammando\orcidlink{0000-0001-7618-7527}}  
\author[1]{F.~De~Gasperin\orcidlink{0000-0003-4439-2627}}  
\author[8]{I.~Delvecchio\orcidlink{0000-0001-8706-2252}} 
\author[8]{F.~Vito\orcidlink{0000-0003-0680-9305}}
\author[17]{K.~Rubinur\orcidlink{0000-0001-5574-5104}}
\author[5]{M.~Krezinger\orcidlink{0000-0002-8813-4884}}
\author[1]{R.~Lico\orcidlink{0000-0001-7361-2460}}
\author[18]{E.~Momjian\orcidlink{0000-0003-3168-5922}}   
\author[1]{M.~Orienti\orcidlink{0000-0003-4470-7094}} 
\author[19]{J.~Hodgson}
\author[1]{C.~Stanghellini\orcidlink{0000-0002-6415-854X}} 
\author[20]{A.~Wang} 
\author[21]{M.~Latif\orcidlink{0000-0003-2480-0988}}
\author[22]{H.~Cao\orcidlink{0000-0003-1514-881X}}

\affiliation[1]{INAF $-$ Istituto di Radioastronomia, Via Gobetti 101, I$-$40129, Bologna, Italy}
\emailAdd{cristiana.spingola@inaf.it}

\affiliation[2]{Institute of Space Sciences (ICE, CSIC), Campus UAB, Carrer de Magrans, 08193, Barcelona, Spain}
\emailAdd{marmezcua.astro@gmail.com}

\affiliation[3]{Institut d'Estudis Espacials de Catalunya (IEEC),  Edifici RDIT, Campus UPC, 08860 Castelldefels, Barcelona, Spain}
            
\affiliation[4]{Shanghai Astronomical Observatory, Chinese Academy of Sciences, 80 Nandan Road, Shanghai 200030, China}
\emailAdd{yuanqi@shao.ac.cn, antao@shao.ac.cn}

\affiliation[5]{Konkoly Observatory, HUN-REN CSFK, MTA Centre of Excellence, Konkoly Thege Mikl\'os \'ut 15-17, Budapest, Hungary}
\emailAdd{frey.sandor@csfk.org}

\affiliation[6]{Institute of Physics and Astronomy, ELTE E\"otv\"os Lor\'and University, P\'azm\'any P\'eter s\'et\'any 1/A, Budapest, Hungary}

\affiliation[7]{Max-Planck-Institut f\"ur Astronomie, K\"onigstuhl 17, D-69117, Heidelberg, Germany}
\emailAdd{belladitta@mpia.de}

\affiliation[8]{INAF -- Osservatorio di Astrofisica e Scienza dello Spazio, via Gobetti 93/3, I-40129, Bologna, Italy}

\affiliation[9]{INAF - Osservatorio Astronomico di Brera, via Brera 28, $I-$20121 Milan, Italy}

\affiliation[10]{Center for Astrophysics—Harvard \& Smithsonian, 60 Garden Street, Cambridge, MA 02138, USA}

\affiliation[11]{Dipartimento di Fisica e Astronomia “Augusto Righi”, Alma Mater Studiorum Università di Bologna, Via Gobetti 93/2, $I-$40129
Bologna, Italy}

\affiliation[12]{Joint Institute for VLBI ERIC (JIVE), Oude Hoogeveensedijk 4, 7991 PD Dwingeloo, The Netherlands}

\affiliation[13]{Faculty of Aerospace Engineering, Delft University of Technology, Kluyverweg 1, 2629 HS Delft, The Netherlands}

\affiliation[14]{Department of Astronomy, Institute of Physics and Astronomy, ELTE E\"otv\"os Lor\'and University, P\'azm\'any P\'eter s\'et\'any 1/A, Budapest, Hungary}

\affiliation[15]{HUN-REN–-ELTE Extragalactic Astrophysics Research Group, ELTE E\"otv\"os Lor\'and University, P\'azm\'any P\'eter s\'et\'any 1/A, Budapest, Hungary}

\affiliation[16]{Institute of Astronomy, Faculty of Physics, Astronomy and Informatics, Nicolaus Copernicus University, Grudzi\c{a}dzka 5, 87-100 Toru\'n, Poland}

\affiliation[17]{Institute of Theoretical Astrophysics, University of Oslo, P.O. Box 1029, Blindern, 0315 Oslo, Norway}

\affiliation[18]{National Radio Astronomy Observatory, 1011 Lopezville Rd., Socorro, NM 87801, USA}

\affiliation[20]{Institute of High Energy Physics, Chinese Academy of Sciences, 19A Yuquan Road, Beijing 100049, China}

\affiliation[21]{Physics Department, College of Science, United Arab Emirates University, PO Box 15551, Al-Ain, UAE}

\affiliation[22]{Shangqiu Normal University, Shangqiu, China}

\affiliation[23]{Department of Astronomy, University of Science and Technology of China, Hefei, Anhui 230026, P.R. China}

\abstract{Active Galactic Nuclei (AGN) are among the brightest sources in the Universe, and those that are also \emph{jetted} are uniquely valuable at the earliest epochs, because their relativistic outflows can regulate the gas supply of their host galaxies, potentially affecting both early star formation and the rapid growth of supermassive black holes (SMBHs). Their compact, high-brightness-temperature radio cores provide the sharpest beacons for very long baseline interferometry (VLBI), enabling direct constraints on Doppler boosting, jet duty cycles, and jet--environment coupling at extreme redshifts. 
In this White Paper, we discuss how the SKA-VLBI will provide sub-$\mu$Jy sensitivity together with milliarcsecond (mas) angular resolution to image and characterise jetted AGN at $z>6$ across SKA-Mid and SKA-Low frequencies. These observations can directly test SMBHs ($>10^6$ M$_{\odot}$) formation/evolution models (including jet-assisted super-Eddington phases) and infer the geometry of the Universe, directly probing the cosmological framework at high precision. Synergies with current and next-generation multi-band facilities will also be crucial to fully understand their host galaxies and their environment, providing an unprecedented panchromatic knowledge of the first jetted AGN.}

\begin{document}
\maketitle

% CS
\section{Introduction}
%Cristiana Spingola

Little is currently known about the first jets from AGN, but they could offer direct insights into the early Universe, particularly in understanding the extreme growth of SMBHs within the first Gyr.

Explaining this rapid growth remains one of the biggest open questions in current studies of SMBH formation and evolution, and jet-assisted accretion has been proposed as a possibility for the growth of SMBHs up to masses of $\sim10^{10}$ M$_{\odot}$ in less than a billion years \citep{jolley2008,jolley2009,ghisellini2013,piana2021,volonteri2021,connor2024,piana2024,mazzucchelli2025,walter2025}.

AGN jets are also expected to play a crucial role in shaping their host galaxies by regulating the rate at which stars are formed across several physical scales, determining the subsequent co-evolution with the host galaxy \citep{gaibler2012,morganti2013,silk2013,murthy2022, mazzucchelli2025}. Since jets propagate at relativistic velocities, they are able to transport energy from the central parts to the outskirts of their host galaxies, also at high-$z$ \citep{saikia2022, kappes2022, gloudemans2025_jet}. 
At $z\gtrsim6$, jets propagate through denser and more gas-rich interstellar medium (ISM)/circumgalactic medium (CGM), while the cosmic microwave background (CMB) energy density increases as $(1+z)^4$, potentially enhancing inverse-Compton cooling, reducing the lifetime and altering jet/lobe visibility. Systematic differences in morphology and spectral curvature relative to low-$z$ jets are, therefore, expected.
Additionally, radio-powerful jetted AGN at $z>6$ serve as unique background sources for probing neutral hydrogen absorption during Epoch of Reionization \cite[EoR,][]{carilli2004}, inferring the physical conditions in the early Universe.  Moreover, they represent beacons for finding distant massive galaxies and protoclusters \cite[e.g.,][]{orsi2016}.

Once the radio emission from these sources is detected by exploiting the unprecedented sensitivity of SKA wide-field surveys, follow-up observations at (much) higher angular resolution will be crucial to spatially resolve the jet structure at high-$z$. 
Indeed, radio observations at mas and sub-mas angular resolution with VLBI have been extremely valuable in quantifying the physical conditions within AGN jets \cite[e.g.,][]{frey2003, frey2010, perger2019, spingola2020, momjian2021, koller2025, gloudemans2025_jet} and in inferring cosmological parameters \citep{roland1993, kellerman1993, stelmach1994, gurvits1994, gurvits1999, hodgson2020, an2020}. 
At the same time, the much smaller synthesized beam of VLBI reduces classical source confusion and mitigates surface-brightness dilution, allowing extremely faint compact cores to be isolated more robustly than in lower-resolution imaging \cite[e.g.,][]{krezinger2024} and enabling a cleaner physical separation between AGN-powered and star-formation-powered radio emission (see \citealt{panessa2019}, for a review).

VLBI observations can, therefore, provide essential information to potentially answer several open questions, as for instance: how can SMBHs accrete mass so fast at such early epochs? How do jets influence the growth of their host galaxies?  How prevalent are jets in high$-z$ AGN, and can we observe their size evolution? Are high-$z$ AGN jets different from local jets once the denser early-Universe environment and enhanced CMB losses are accounted for? Each discovery of a new high-$z$ jetted AGN observed with VLBI provides a step forward to clearly answer these questions. 

In this context, an ideal class for VLBI observations of jetted AGN is that of \textsl{blazars}, i.e. jetted AGN pointing towards us. They are the brightest (hence easier to detect at large distances and with sparser arrays) and can help to derive an unbiased estimate of the space density of the entire class of jetted AGN, becoming a key sample to study the high-$z$ population \cite[e.g.,][]{volonteri2011, ghisellini2014, sbarrato2015, caccianiga2019, banados2025}.

In this White Chapter we give an overview of the remarkable contribution that VLBI has provided in the field of jetted AGN and an outlook on what SKA--VLBI will enable. When used in VLBI mode, SKA-AA4 as a phased-up array will serve as a highly sensitive element in global networks, vastly enhancing the baseline sensitivity and $(u,v)$ coverage.  
SKA-Mid wide-field surveys at unparalleled sensitivity ($\sim 50$~nJy off-source root-mean-square in deep integrations) and angular resolution (sub-arcsecond at GHz frequencies) enable the direct detection of synchrotron emission from relativistic jets and lobes associated with SMBHs at $z > 6$ (e.g., \citealt{godfrey2017}, Sec. 2, 3 and 4). The broad frequency coverage of SKA-Low and SKA-Mid (from 50 MHz to 15 GHz) eases spectral studies and polarization measurements, yielding directly magnetic field strengths, electron energy distributions, and jet dynamics \citep{mingo2022}. 
With the intercontinental baselines, SKA--VLBI will spatially resolve jet morphology on scales up to few kpc at high image fidelity, critical for studying jet collimation, bending, and interaction with the ambient intergalactic medium (IGM) during the EoR.

This chapter is structured as follows: Sections \ref{sec:2} and \ref{sec:3} describe the formation models of the first SMBHs and the current proposed theories for their fast accretion. Section \ref{sec:4} is devoted to the state-of-the-art techniques for finding jetted AGN up to the largest distances. Sections \ref{sec:5} and \ref{sec:6} discuss the VLBI role in assessing the physical properties of high-$z$ jets and testing the current cosmological model.  The role of very high angular resolution observations, enabled by VLBI, to better understand AGN feedback is reported in Section \ref{sec:7}. We conclude by highlighting the synergies of SKA--VLBI with current and the next generation multi-wavelength facilities in Section  \ref{sec:8}. 
Throughout this Chapter, we assume $H_0$ = 67.8 km s$^{-1}$ Mpc$^{-1}$, $\Omega_M = 0.31$, and $\Omega_{\Lambda} = 0.69$ \citep{planckcollab2016}; the synchrotron spectral index is defined as $S_{\nu} \propto \nu^{-\alpha}$, unless otherwise stated.

% MM
\section{First black holes from the dark ages to the epoch of reionization}\label{sec:2}
%Mar Mezcua \& her team

SMBHs with masses exceeding a million times that of the Sun ($> 10^6$ M$_{\odot}$) are believed to reside at the centers of all massive galaxies (stellar mass $M_\mathrm{*} > 10^{10}$ M$_{\odot}$) and to play a pivotal role in galaxy formation and evolution. However, the processes by which these SMBHs form and grow remain poorly understood.

The finding of SMBHs as massive as $\sim10^{10}$ M$_{\odot}$ at redshifts $z\sim 6-7$ ($\sim$700 Myr after the Big Bang; \citealt{fan2023}) has long challenged theoretical models. More recently, observations from the James Webb Space Telescope (\textit{JWST}) have pushed these limits further, revealing SMBHs with masses of $10^{6}-10^{7}$ M$_{\odot}$ at even earlier cosmic times ($z$ = 8.7, \citealt{larson2023}; $z$ = 10.3, \citealt{bogdan2024}; $z$ = 10.6, \citealt{maiolino2024}). To attain such masses so early likely requires BH formation as far back as $z \sim$ 20 \cite[see reviews by][]{mezcua2017,greene2020,pacucci2022}.

The main scenarios proposed for the formation of the initial ``seed" BHs are light seeds of $\sim$100 M$_{\odot}$ arising from the collapse of the first generation of Population III stars, and heavy seeds of $\sim10^{4}-10^{5}$ M$_{\odot}$ formed through direct pre-galactic gas collapse (\citealt{loeb1994,lodato2006}) or runaway stellar mergers in dense star clusters \citep{portegies1999,devecchi2009}. Other possibilities include primordial BHs \citep{capelluti2022,ziegler2021} and episodes of super-Eddington accretion \citep{Volonteri2005}, which may also contribute to their rapid early growth.

Because of the challenges of detecting seed BHs at $z \sim$ 20, many studies focused on searching for the local relics of those seed BHs that did not grow nor became supermassive. These relics should be found as BHs of $\sim10^{2}-10^{5}$ M$_{\odot}$ (also called \textsl{intermediate-mass} BHs) and be primarily located in dwarf galaxies, which because of their low mass, low metallicity, and quiet merger history are thought to resemble primordial galaxies.

Hundreds of such intermediate-mass BHs residing in dwarf galaxies in the local Universe have been found to be actively accreting, with inferred masses $M_\mathrm{BH}\sim10^{4}-10^{6}$ M$_{\odot}$ \citep{reines2013, moran2014, marleau2017, chilingarian2018, mezcua2020, lemmingsIII,salehirad2022, mezcua2024, lemmings2024} and out to $z \sim$ 2 \citep{mezcua2016,mezcua2018, mezcua2019, mezcua2023, mezcua2024_apj}.
While most of these findings have made use of optical, infrared, and X-ray diagnostics, a few tens of dwarf galaxies host AGN that present also jet radio emission 
\citep{mezcua2019,reines2020,yang2020,yang2023,wu2024, yuan2025}. Some of these jets are spatially resolved \cite[e.g.,][]{mezcua2018_jet, yang2020} and often present powers as high as jets in massive galaxies \cite[e.g.,][]{mezcua2018, yang2020, yang2023, flores2025}. 
Radio-mode AGN feedback could thus be as significant in dwarf galaxies as in more massive galaxies, which has important implications for models of seed BH and galaxy evolution, as local dwarf galaxy BHs may not be pristine relics of the original seed population because of this feedback. This affects how reliably BH occupation fractions, IMBH demographics, and low-mass BH–galaxy scaling relations can be used to discriminate between light- and heavy-seed scenarios \cite[e.g.,][]{mezcua2019,mezcua2019NatAs}.

Because dwarf galaxies have shallow gravitational potentials, even modest jet powers can have an outsized impact: jets can heat and/or expel a significant fraction of the ISM, regulate subsequent gas inflow and SMBH fueling, and thereby imprint the occupation fraction and duty-cycle of seed BHs. Conversely, jet-driven over-pressurisation may also trigger localised star formation in dense clouds. Quantifying how often jets occur in this low-mass regime, therefore, feeds back directly into seed-formation and early-growth models \cite[e.g.,][]{mezcua2019NatAs}.

\begin{figure}[b!]
\centering
\includegraphics[width=0.7\linewidth]{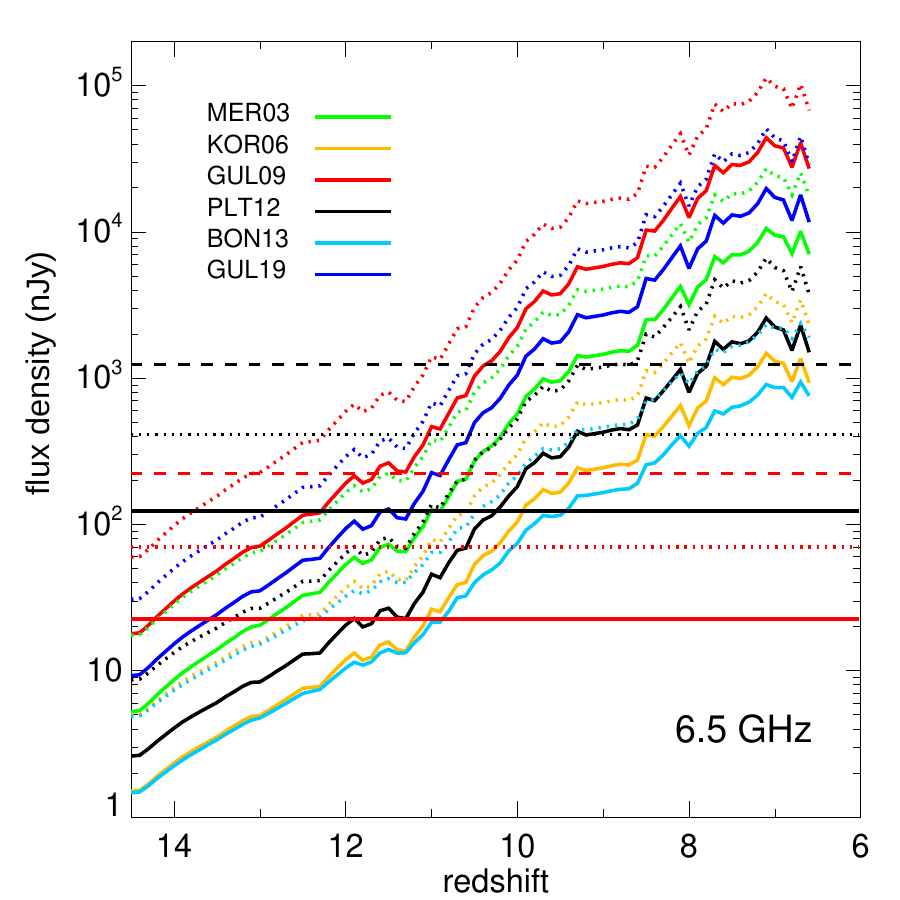}
\caption{BH radio flux densities estimated from the fundamental plane of BH accretion (\citealt{merloni2003}, MER03; \citealt{kolding2006}, KOR06; \citealt{gultekin2009}, GUL09; \citealt{plotkin2012}, PLT12; \citealt{bonchi2013}, BON13; \citealt{gultekin2019}, GUL19) for a spectral index $\alpha$ = 0.7 (solid) and $\alpha$ = 0.3 (dotted) from z = 6–14.5 at SKA-MID bands 1 (500 MHz), 2 (1.5 GHz), 5a (2.5 GHz), and 5b (6.5 GHz). The dashed, dotted, and solid
horizontal lines are the detection limits for 1 h, 10 h, and 100 h integration times for SKA-MID (black) and ngVLA (red). Figure and caption adapted from \cite{latif2024}.}
\label{fig:Latif2023}
\end{figure}

The \textit{JWST} is discovering an increasing number of SMBHs at early epochs, many of which are dust-reddened and located in low-mass galaxies, which suggests that the first BHs accrete at high rates and that their growth is highly obscured by dust \cite[e.g.,][]{matthee2024, greene2024}. 
The SKA will be pivotal to circumvent dust obscuration, identify BHs growing at low rates, and reveal extended features  that could be indicative of radio jets \cite[e.g.,][]{yue2021}. To unravel the role of large scale jets in the BH growing phase, spatially resolved VLBI observations are necessary. A seed BH of $M_\mathrm{BH} \sim 10^{6}$ M$_{\odot}$ and flux density $\sim$100 nJy could be detected with the SKA-Mid out to $z \sim 15$ with an integration time of 100 hours (see Fig.~\ref{fig:Latif2023}; \citealt{whalen2021}), probing the evolution at later times of these objects into the so-called Little Red Dots (LRDs; \citealt{latif2025}) or quasars of $M_\mathrm{BH} \sim 10^9$ M$_{\odot}$ \citep{latif2024}.

We estimate the radio flux density at 6.5 GHz based on the fundamental planes (FPs) of BH accretion, which are empirical relationships between BH mass, $M_\mathrm{BH}$, nuclear radio luminosity at 5 GHz, $L_\mathrm{R}$, and nuclear X-ray luminosity at 2 - 10 keV, $L_\mathrm{X}$ \citep{merloni2003}.  We determine $L_\mathrm{X}$ from the bolometric luminosity $L_{\mathrm{bol}}$, with Equation 21 of \citet{marc04}:
\begin{equation}
\mathrm{log}\left(\frac{L_\mathrm{bol}}{L_\mathrm{X}}\right) = 1.54 + 0.24 \mathcal{L} + 0.012 \mathcal{L}^2 - 0.0015 \mathcal{L}^3,
\end{equation}

where $L_\mathrm{bol}$ is in solar luminosity units and $\mathcal{L} = \mathrm{log} \, L_\mathrm{bol} - 12$.   $L_\mathrm{R}$ in the source frame can be calculated from $L_\mathrm{X}$ from FP of the form:

\begin{equation}
\mathrm{log} \, L_\mathrm{R} (\mathrm{erg \, s^{-1}})= A \, \mathrm{log} \, L_\mathrm{X} (\mathrm{erg \, s^{-1}}) + B \, \mathrm{log} \, M_\mathrm{BH} (\mathrm{M}_{\odot})+ C,
\end{equation}

where $A$, $B$, and $C$ are taken from \citet{merloni2003}, \citet{kolding2006}, \citet{gultekin2009}, \citet{plotkin2012}, and \citet{bonchi2013}.

The radio flux density from high$-z$ sources is  redshifted into a given receiver band today and does not come from 5 GHz in the source frame. Therefore, we calculate the source frame flux from $L_\mathrm{R} =$ $\nu L_{\nu}$, assuming that the spectral luminosity $L_{\nu} \propto \nu^{-\alpha}$.  We consider $\alpha =$ 0.7 and 0.3 to enclose a reasonable range of spectral profiles \citep{ccb02,glou21}.  The spectral flux at $\nu$ in the observer frame is then determined from the spectral luminosity at $\nu'$ in the rest frame from
\begin{equation}
F_\nu = \frac{L_{\nu'}(1 + z)}{4 \pi {d_\mathrm L}^2},
\end{equation}
where $\nu' = (1+z) \nu$ and $d_\mathrm L$ is the luminosity distance.

% TA
\section{SMBH accretion beyond the Eddington limit }\label{sec:3}
% Tao An \& his team

Even with the most massive stellar seeds, canonical thin-disc accretion at the Eddington rate fails to account for the rapid growth of SMBHs within less than a billion years \citep{Volonteri2005, inayoshi2020}. This discrepancy implies that super-Eddington accretion phases are unavoidable in explaining the formation of the first SMBHs. 
A key point for this White Chapter is that the most extreme growth channels have concrete radio-VLBI signatures: high-$T_{\rm b}$ compact cores, spectral turnovers from synchrotron self-absorption or free--free absorption, variability that constrains Doppler factors, and (with multi-epoch imaging) structural evolution that constrains jet speeds and duty cycles. SKA--VLBI can therefore test super-Eddington/jet-assisted growth by directly measuring these observables for statistically meaningful samples at $z\gtrsim6$.

Relativistic jets are expected to facilitate super-Eddington accretion by removing angular momentum from the accretion disk, even when radiation pressure approaches isotropy, allowing accretion rates beyond the Eddington limit. Semi-analytic models that incorporate a ``jet-enhanced'' accretion suggest that brief super-Eddington bursts ($\lambda \equiv L/L_{\rm Edd} \approx 5$--$20$) embedded in longer sub-Eddington intervals over $\sim10$--30 Myr can reproduce the $z \gtrsim 6$ SMBH mass function \citep{jolley2008, piana2021, volonteri2021}.

Recent discoveries offer direct observational evidence supporting this theoretical framework. For instance, VLASS J0410$-$0139, a blazar at $z \approx 7$, is powered by a $7 \times 10^8 \, M_\odot$ SMBH with $\lambda \approx 1.2$ after accounting for relativistic beaming effects \citep{banados2025}.  Beaming statistics imply a much higher parent population of misaligned systems, but the inferred space densities remain sensitive to uncertainties in $M_{\rm BH}$, $\lambda_{\rm Edd}$, and selection functions at the highest redshifts. The current evidence is, therefore, best phrased as \emph{consistent with} potentially widespread jet-assisted growth during the EoR rather than as a definitive proof.

These jets play a crucial role in shaping their environments through lobe expansion that preheats protogalactic gas, modifies star formation across kiloparsec scales, and seeds magnetic fields. All these consequently influence galaxy formation \citep{gaibler2012, ruszkowski2009}, requiring detailed observations of both jet launching mechanisms and environmental impact.

Recent radio studies of super-Eddington AGN provide key benchmarks for high-redshift systems. Narrow-line Seyfert 1 galaxies are often considered prime laboratories for testing extreme accretion physics, and VLBI observations of Mrk 110 \citep{wang2025} revealed episodic superluminal knots ($\beta_{\rm app} \approx 3.6c$) emerging from a radio-quiet nucleus, appearing only during high optical and X-ray states. These observations are consistent with magnetically arrested disk (MAD) configurations and demonstrate that transient accretion states can lead to relativistic jets even in radio-quiet AGN. Notably, Mrk~110 exhibits striking radio spectral evolution from steep ($\alpha \approx 0.63$) to inverted ($\alpha \approx -0.69$) as the jet flux density doubled, with the jet later decelerating to $\sim 1.5c$ at $\sim 1.1$ pc near the transition between broad-line and narrow-line regions. This behavior aligns with MAD-based interpretations for state-dependent jet launching, and motivates the view that the observed radio-loud/quiet dichotomy can be affected by temporal variability, with intermittent jet activity linked to changing accretion states rather than representing a permanently fixed classification (e.g., \citealt{an2026timedomain}).

Systematic VLBI surveys of Palomar--Green quasars reinforce the ubiquity of compact radio cores in radio-quiet AGN \citep{wang2023a, wang2023b, chen2025}. \citet{wang2023a} detected VLBI cores in 10 of 16 RQ quasars, implying that weak jet activity is more prevalent than previously thought. Moreover, multi-frequency monitoring has revealed radio flares with spectral flattening, likely resulting from magnetic reconnection near jet bases \citep{berton2018, chen2020}, establishing scaling relations between brightness temperature, Eddington ratio, and BH mass. Interestingly, observations of dwarf galaxies hosting intermediate-mass BHs also show compact radio cores with $T_{\rm b} > 10^8$ K, confirming that relativistic jets can emerge across the BH mass spectrum, even during super-Eddington phases \citep{greene2020}. These findings support the theory that seed BHs in the early Universe likely experience rapid, jet-assisted growth.

Despite advances, identifying super-Eddington accretors at $z > 6$ remains a challenging observational problem \cite[e.g.,][]{wolf2023,ighina2025_superE}. Virial BH mass estimates based on single-epoch spectroscopy often suffer from systematic uncertainties, while heavy dust obscuration can mimic signatures of high Eddington ratios. Nevertheless, several diagnostic flags such as extreme Fe II emission, weak [O III] lines, steep UV continua, and soft X-ray excesses can help identify super-Eddington candidates at high$-z$. Radio observations, in particular, are becoming increasingly important. 
At the bright end of current $z\sim6$ radio samples, the implied radio powers are often too large to be explained by star formation alone for typical host galaxies, making an AGN origin more plausible; however, the angular resolution of wide-area surveys is generally insufficient to morphologically separate compact star formation and AGN components at high redshift, so VLBI-determined brightness temperatures and core/jet morphologies remain the decisive discriminator.
In local AGN samples, radio loudness shows trends with Eddington ratio, especially towards the radio-loud tail, but with substantial scatter and selection dependence; the relationship becomes more complex around $\lambda_{\rm Edd}\sim1$ due to radiative feedback and geometry \citep{macfarlane2021, Yue2024_lrd}.
Crucially, we do \emph{not} advocate inferring $\lambda_{\rm Edd}$ from radio data alone at high redshift: rather, radio-VLBI measurements provide complementary constraints on jet power, Doppler boosting, and duty cycle that can be combined with optical/infrared/X-ray information to tighten the overall growth picture.

The SKA-VLBI will dramatically enhance our understanding of super-Eddington SMBH accretion at high redshift by providing sub-$\mu$Jy sensitivity and an angular resolution $\lesssim 0.8$~mas up to Band 5b (at a maximum observing frequency of 15.4~GHz). This angular resolution and frequency band could be exploited also with precursors (such as MeerKAT-VLBI) in the foreseeable future\footnote{\href{https://inaf.ubuy.cineca.it/PortaleAppalti/do/FrontEnd/DocDig/downloadDocumentoPubblico.action;jsessionid=DE1A5074086A522F9230C16A248181B2.ubuy-inaf-fo-prod-1-ubuybe38?codice=G01172&id=10030&idprg=&_csrf=U34CE4VX17SIS70L9WAFMVEGCO6SLRF}{Italian National Institute for Astrophysics (INAF) Document "MeerKAT Band 5b Receivers
Technical Specifications".}}. At $z\simeq6$, $0.8$ mas corresponds to only a few parsecs, so SKA--VLBI will directly access the innermost jet-launching region at cm-wavelengths.
These observations will enable direct imaging of the jet bases in $z > 6$ quasars, allowing measurement of Doppler factors and separation of relativistic beaming from intrinsic luminosity. Multi-epoch constraints on proper motions at $z\gtrsim6$ will generally require time baselines of at least $\sim5$--10 yr (and often longer, depending on component identification and signal-to-noise (S/N) ratio; see Sect.~\ref{sec:6}), so starting long-term monitoring early is essential. 
The SKA’s broad frequency coverage will further enable spectral and absorption studies, revealing how early jets interact with their environments during the EoR. Coordinated observations with the Atacama Large Millimeter Array (ALMA) and \textit{JWST} will link jet activity to host galaxy properties, enabling systematic studies of super-Eddington accretion across cosmic time (see Sec. \ref{sec:8}).
These data will illuminate the feedback mechanisms and growth cycles of the first SMBHs, bridging the gap between theoretical models and the observed rapid emergence of $>10^9$~$M_\odot$ quasars at $z \sim 6$–10.

% SB
\section{Finding AGN jets at $z \gtrsim 6$}\label{sec:4}
%Silvia Belladitta \& her team

\begin{figure*}
    \centering
    \includegraphics[width=0.9\linewidth]{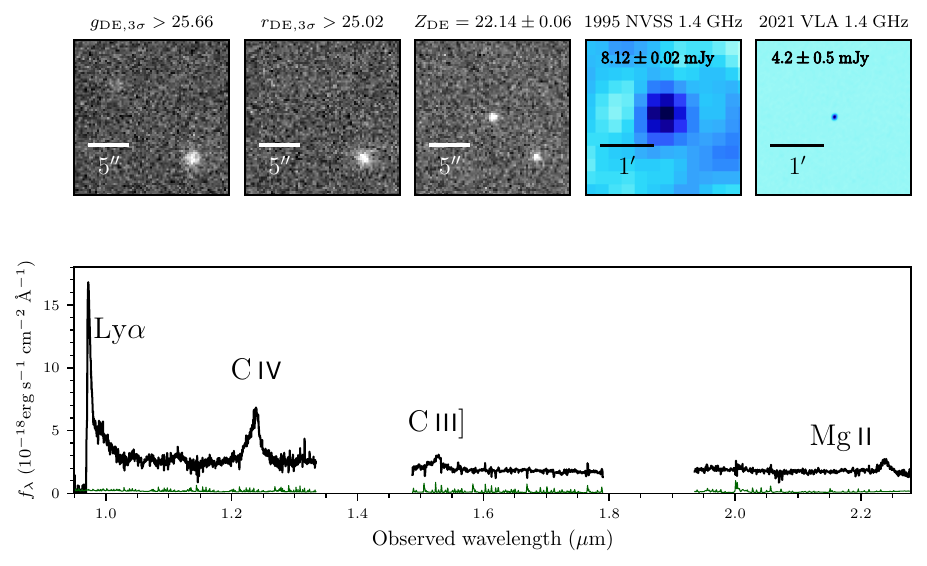}
    \caption{A dropout radio source at $z=7$: photometry and spectrum of the quasar J0410$-$0139. The greyscale postage stamps show optical images, while the radio panels show the 1995 NVSS and the 2021 VLA 1.4~GHz images, both at 1.4~GHz. The 2021 VLA observations at 1.4~GHz, shown at the same size for comparison, confirms the presence of a single radio source in the field with clear evidence of variability (also confirmed by other radio observations in \citealt{banados2025}). The bottom panel presents the optical/NIR spectrum as a combination of VLT/FORS2, Keck/NIRES, Magellan/FIRE, and LBT/LUCI. Adapted from \cite{banados2025}.}
    \label{fig:dropout}
\end{figure*}

High$-z$ jetted AGN selection typically begins with wide-area radio surveys, followed by optical/near-IR cross-matching to identify photometric dropouts (see Fig.~\ref{fig:dropout}). Spectroscopic detection of emission lines and/or the Lyman break provides definitive confirmation of their high$-z$ nature (see Fig.~\ref{fig:dropout}).
For decades, the Northern-hemisphere facilities such as the NRAO VLA Sky Survey (NVSS, \citealt{condon1998}) and the Faint Images of the Radio Sky at Twenty-Centimeters (FIRST, \citealt{becker1994}) survey, at 1.4~GHz, were the main surveys for discovering jetted AGN across cosmic time, including the first jetted AGN at $z>6$ \cite[e.g.,][]{mcgreer2006,frey2011,zeimann2011,banados2015,belladitta2020}.
It was only at the beginning of the 2020s that the number of AGN with radio jets in the EoR increased by a factor of $\sim$4 to over 20 sources \cite[e.g.,][]{banados2021,banados2023,gloudemans2022,ighina2025}, thanks to new wide-area, deep radio surveys, 
such as the LOw-Frequency ARray (LOFAR, a SKA pathfinder) Two-metre Sky Survey (LoTSS, \citealt{shimwell2019,shimwell2022}) at 150~MHz, the surveys carried out with SKA precursors (ASKAP), such as the Rapid ASKAP Continuum Survey (RACS, \citealt{mcconnell2020,hale2021,duchesne2025}) at $0.888-1.66$~GHz (which largely extended the discoveries of jetted AGN in the Southern hemisphere), and the Karl G. Jansky Very Large Array Sky Survey (VLASS, \citealt{lacy2020}) at 3~GHz.
These facilities can detect compact AGN candidates down to the $\sim$mJy level.

The \textit{radio$+$optical/IR technique} is optimized for identifying un-obscured (Type I) AGN.
A key advantage of radio selection in this context is that it excludes the main contaminant populations in dropout searches (i.e., stars), as most of them are radio weak or silent.
However, theoretical and observational studies suggest that the majority of high$-z$ SMBHs are obscured ($>80\%$, \citealt{vito2018,gilli2022,CapettiBalmaverde2024}), and 
therefore missed by this approach.
To reveal these obscured or Type II AGN, cross-matching radio data with mid-/far-IR or sub-mm observations offers a complementary strategy \cite[e.g.,][]{truebenbach2017}, exploiting the ability of radio waves to penetrate dense gas and dust due to their low  opacity at these frequencies  \citep{hildebrand1983}.
This method has proven effective in detecting obscured radio AGN at $z>6$. 
By combining radio observations at $0.144 - 3$~GHz with mid- to far-IR and sub-mm ($24-850\,\mu$m) data, \cite{endsley2022} identified a radio AGN candidate, which was later spectroscopically confirmed at $z=6.83$ through the detection of the [C II] $158\,\mu$m emission line (\citealt{endsley2023}).

\begin{figure*}
\centering
    \includegraphics[width=0.95\linewidth]{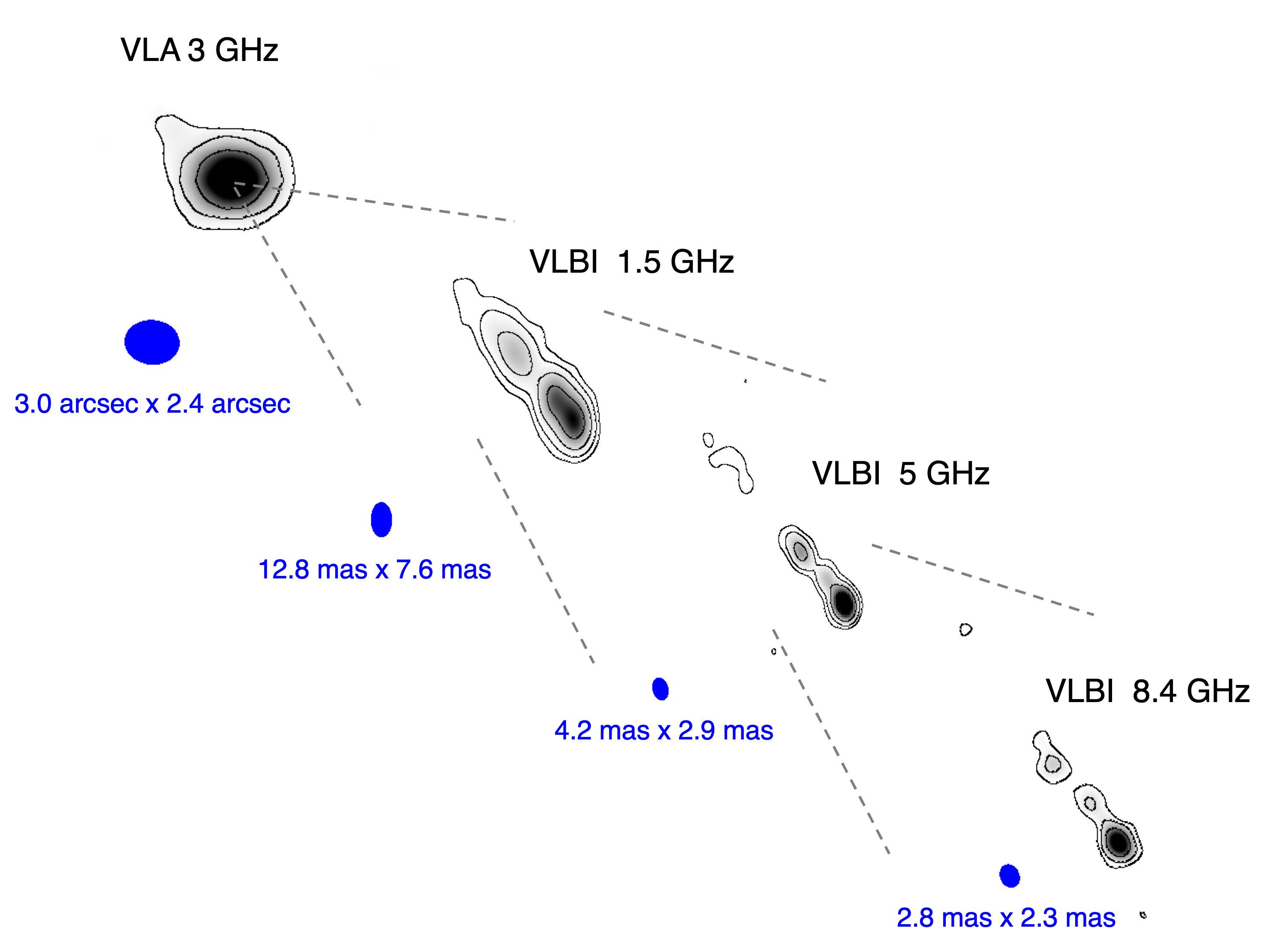}
    \caption{Example of how VLBI can spatially resolve the jet structure in the source PSO J0309+27 at $z=6.1$ (adapted from \citealt{belladitta2020} and \citealt{spingola2020}). The restoring beam relative to each observation and the observing frequencies are reported directly in the figure. The three VLBI bands shown here correspond to the SKA Band 2, Band 5a and Band 5b.}\label{fig:vlbi_resolving_jet}

\end{figure*}

High$-z$ jetted AGN candidates selection can rely on radio spectral indices (flat, steep, or inverted spectra) and variability to identify AGN-like behavior.
Interpreting spectral-index trends at the highest redshifts requires care, since selection effects and enhanced inverse-Compton losses off the CMB can suppress extended steep-spectrum emission, biasing samples toward core-dominated compact sources (see also Sec.~\ref{sec:6}).
Notable examples of these targeted selections include: 

\noindent \textit{i)} Flat Spectrum Radio Quasars (FSRQs, i.e. blazars). 
Currently, the most distant blazar known is at $z=7$ (\citealt{banados2025}), discovered partly through rapid multi-frequency radio variability (see Fig.~\ref{fig:dropout}), with only one other known beyond $z>6$ (see \citealt{belladitta2020}, Fig.~\ref{fig:vlbi_resolving_jet}). Their existence implies that there must be hundreds of jetted AGN at $z>6$ that still need to be discovered in the same area and at the same optical limit. Hence, blazars are ideal sources to enable a complete census of the jetted AGN population across cosmic time. 

\noindent \textit{ii)} Powerful High$-z$ Radio Galaxies (HzRGs), expected to dominate the population (e.g., \citealt{miley2008}), identified mainly via Ultra Steep Spectrum (USS) selection or low-frequency spectral curvature (e.g., \citealt{saxena2018a,drouart2020,broderick2022}), but see also \cite{capetti2025} for a different selection technique based on optical colors. The origin of the USS is complex, as it could be related to the losses due to inverse-Compton with the CMB photons (i.e., stronger spectral steepening at high frequencies); selection effects (i.e., flux-density-limited surveys and spectral index cuts bias samples toward steeper spectrum sources); the K-correction (i.e., observing at fixed frequency bands probes higher rest-frame frequencies at high$-z$, where spectra are intrinsically steeper due to radiative losses); different environments at high$-z$ (i.e., denser environments could, in principle, produce steeper electron energy distributions and thus steeper spectra; see \citealt{Morabito2018} for a detailed discussion).
The most distant radio galaxies discovered so far have redshifts $\sim5$ (\citealt{vanBreugel1999,drouart2020}).
These powerful HzRGs trace the most massive dark matter overdensities and host the most massive BHs in the early Universe (e.g., \citealt{mayo2012,drouart2014}). 

\begin{figure*}
    \centering
    \includegraphics[width=0.7\linewidth]{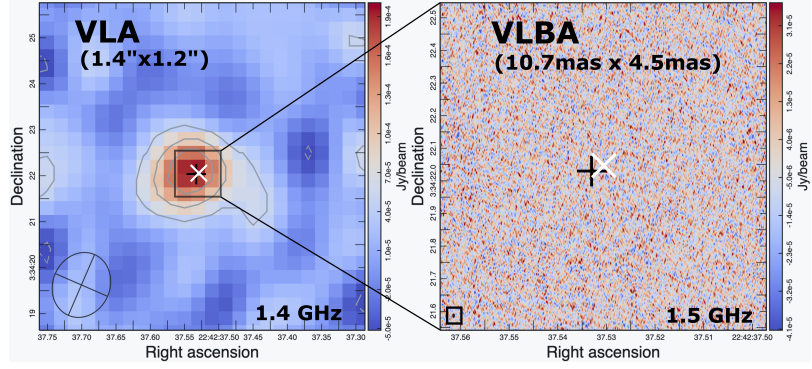}
    \caption{The role of VLBI observations in disentangling AGN radio emission. \textit{Left:} VLA map at 1.4~GHz for the quasar J2242$+$0334. The contour levels are at $[-1, 2, 4, 6]$ times the RMS noise ($24.7\,\mu$Jy/beam). \textit{Right:}  $1^{\prime\prime} \times 1^{\prime\prime}$ VLBA image at 1.5~GHz, zooming into the central area of the VLA map.  
    In both images, the synthesized beam is shown at the bottom left corner. For both panels, the black plus sign denotes the optical position of the source and the white cross represents the peak position of the 3-GHz VLA observations reported in \cite{liu2021}. Adapted from \cite{liu2022}.}
    \label{fig:vlavlba}
\end{figure*}

The radio-excess method, which relies on using mid-IR colors and radio luminosity (\citealt{hardcastle2025}), can distinguish AGN from star-forming galaxies, but separating jet from non-jet emission remains challenging.
The VLBI technique is essential for this distinction, providing mas-scale resolution to analyze morphology, brightness temperature ($T_\mathrm{b}$), and spectral index (e.g., \citealt{lister2019}, but see also \citealt{lobanov2015} for a visibility-based approach for estimating $T_\mathrm{b}$).
Jet-driven emission typically shows core/core–jet collimated structures, with high brightness temperatures ($T_\mathrm{b}>10^6$\,K) and flat-to-steep ($\alpha= 0.22-1.1$) spectral indices (e.g., \citealt{kovalev2005,hovatta2014,lister2019}). 
In contrast, AGN winds/outflows and intense star formation  tend to produce more diffuse, low-$T_\mathrm{b}$ emission with steeper spectra, while coronal emission appears as compact, low-$T_\mathrm{b}$ sources (e.g., \citealt{condon2013,inoue2018,panessa2019}). 
Additional indicators like polarization, variability, and jet component proper motion provide further discrimination (e.g., \citealt{jorstad2005,lister2019}).
The VLBI approach is particularly important for low-power jetted radio sources ($L_{1.4\rm GHz}\leq10^{23}$ W~Hz$^{-1}$, e.g., \citealt{panessa2019, Baldi2023_reviewFR0}), where the jet emission can be comparable to star formation or outflows and winds. 
However, it is also applicable to powerful jetted objects.
The quasar J2242$+$0334 at $z=5.9$ (\citealt{liu2022}, see Fig.~\ref{fig:vlavlba}) 
exemplifies this challenge: it shows compact radio morphology on kpc scales (VLA-detected) but lacks a VLBI counterpart at pc scales. This absence suggests either weak, un-collimated jets below VLBA detection thresholds or alternative AGN emission mechanisms (e.g., strong winds), highlighting VLBI's critical role in verifying jet activity.

All the selection methods outlined above will be applicable —and greatly enhanced— in the SKA era.
Thanks to its unprecedented sensitivity, wide sky coverage, and broad frequency range, SKA-Mid and SKA-Low will be game changers in the identification of jetted AGN during the EoR. 
For instance, based on the number of high$-z$ blazars observed at $\sim$mJy levels, \cite{caccianiga2024} predicts over 1000 Type-I jetted AGN at $z>4$ detectable in the SKA-Mid over an area of $\sim$10000~deg$^2$ down to 100~$\mu$Jy/beam and magnitude 21.
Moreover, despite uncertainties related to the unknown radio luminosity function in the early Universe, \citet{mazzolari2024} estimate that $\sim$2000 radio AGN at $z>6$ and $\sim$30 at $z>10$ are detectable in the SKA-Mid Wide survey (over an area of 1000~deg$^2$, \citealt{prandoni2015}) with a 5~$\mu$Jy detection limit, including sources whose radio emission may not originate from jets (but from e.g., star formation or AGN winds). 
The estimates of \citet{mazzolari2024} do not consider radio losses for CMB quenching or synchrotron self-absorption (SSA). \cite{afonso2015} report that taking into account CMB losses, in an area of $1000-5000$ deg$^2$ at a detection level of 10~$\mu$Jy, SKA surveys at 1 GHz can detect $\sim$50000, 20000, and 5000 jetted AGN at $z>6,8,$ and $10$, respectively. These estimates are very optimistic and  different from those reported in \citet{mazzolari2024}. One of the main reasons is the different treatment of the radio luminosity function and the assumption on the radio spectral index. 
However, despite these discrepancies, both studies agree on a key point: the SKA will detect thousands of radio sources in the EoR. The dominant bottleneck will be identifying which of these radio sources truly lie at $z\gtrsim6$, requiring efficient multi-wavelength counterparting and ultimately spectroscopic confirmation.

SKA--VLBI (using SKA as a \textit{super station} in a global VLBI network) will be essential to distinguish jetted from non-jetted sources.  
Even during early science phases (SKA-Mid operating at 50\% of its capacity), SKA--VLBI will match or exceed the capabilities of today's interferometers such as European VLBI Network, but with the advantage of accessing the entire southern sky and the galactic center (\citealt{paragi2015}).
With an angular resolution reaching $\sim$0.8 mas at $3-8$~GHz and sensitivity down to a few $\mu$Jy beam$^{-1}$ (\citealt{paragi2015,li2024}), SKA--VLBI will enable detailed imaging of AGN jets at sub-galactic scales at $z>6$, revealing jet bases and structural evolution in high$-z$ AGN (see also Fig. \ref{fig:vlbi_resolving_jet} for an example). 
It should be noted that studies with SKA--VLBI will be limited by the small field of view, as conducting large-area surveys with SKA--VLBI faces significant technical and computational challenges (\citealt{paragi2015}). Nonetheless, wide-field surveys with SKA--VLBI remain a key priority for the VLBI community, and considerable effort is being invested to explore their feasibility.

Since VLBI observations are necessary to disentangle jetted from non-jetted sources, particularly at low radio luminosities where the jet emission can be comparable to star formation or outflows and winds, we have explored the feasibility of detecting such low-power jetted AGN ($L_{\nu} \approx 10^{23}$\,W\,Hz$^{-1}$) at $z \approx 6$ with an SKA-VLBI array composed by the European VLBI Network (EVN) antennas and SKA AA4 (e.g., Fig. \ref{fig:uvcoverage}).
The required integration time is highly dependent on the spectral index, $\alpha$.  The $1\sigma$ image sensitivity of the EVN + SKA AA4\footnote{Estimated using \url{https://planobs.jive.eu/}} is
$3.8\,\mu$Jy\,hr$^{-1}$, hence for a typical steep-spectrum jet ($\alpha=0.7$), a $5\sigma$ detection at $z=6$ ($z=7$) would require $\gtrsim 600$ ($\gtrsim 1000$) hours. However, for a flat-spectrum source ($\alpha=0.0$), the required observing time decreases to $\approx 40$ ($\approx 60$) hours at $z=6$ ($z=7$), excluding calibration overheads. These results indicate that while mapping extended, low-surface-brightness jet structures remains a challenge for the next generation of deep-field surveys, the detection and identification of compact AGN cores will be well within the reach of targeted SKA-VLBI observations with AA4 already.

From a technical perspective, SKA--VLBI follow-up of faint high$-z$ targets will rely on robust phase calibration strategies (phase referencing, multi-view calibration, and/or in-beam calibrators enabled by SKA beamforming). Reliable phase solutions require S/N $\gtrsim5$--10 per solution interval; with SKA as one of the most sensitive array elements, this pushes the correlated-flux threshold for calibrators into the sub-mJy--mJy regime for minute-scale solutions, depending on bandwidth and partner-station sensitivity. At low frequencies, direction-dependent ionospheric effects and scattering become dominant error terms; at high frequencies, tropospheric phase stability and intrinsic source structure evolution become limiting systematics. These calibration/processing challenges (beyond raw computing) should be considered in the SKA--VLBI readiness roadmap.

A major current limitation in discovering jetted AGN during the EoR is the insufficient depth and coverage of wide-area optical/near-IR photometric surveys (for Type I objects) and sub-mm campaigns (for Type II sources), which are crucial for reliably identifying counterparts to faint radio sources detected by facilities like SKA. 
Without accurate multi-wavelength identifications, many radio detections remain ambiguous, hindering robust population studies and follow-up investigations. 
This challenge is expected to be substantially alleviated with the advent of ongoing and imminent deep, wide-field surveys such as the \textit{Euclid} mission (\citealt{scaramella2022,mellier2024}) and the Legacy Survey of Space and Time (LSST) conducted by the Vera C. Rubin Observatory (\citealt{ivezic2019}). 
In the sub-mm regime, this gap will be covered by the Atacama Large Aperture Submillimeter Telescope (AtLAST, \citealt{Mroczkowski2025}) that with a field of view of $2^{\circ}$ and resolution of few-to-tens arcsec will provide a complete, homogeneous and unbiased picture of dusty sources in the early Universe.

% CS
\section{Inferring physical properties of AGN jets at the earliest epochs with SKA--VLBI}\label{sec:5}
%Cristiana Spingola \& her team

The SKA--VLBI will allow for precise measurements of brightness temperature, magnetic field strength and orientation, and relativistic beaming effects, which are essential for constraining physical parameters such as bulk Lorentz factor ($\Gamma$), jet power, and particle composition.  
Typical values of $\Gamma$ for AGN jets range from a few ($\sim 2 - 5$) up to tens ($\sim 10 - 50$) for blazars. 
Determining $\Gamma$ is crucial for constraining the intrinsic properties of the source, such as its luminosity. 

If multiple visits will be envisioned, SKA--VLBI will be able to resolve transverse jet structures and measure proper motions over time ($\sim$ tens of years timescales at $z>6$), allowing for direct estimates of apparent velocities and hence constraints on $\Gamma$ (especially at higher frequencies, from Band 3 and above). 
By combining $\Gamma$ with the flux density measurements, one can break degeneracies between orientation effects and intrinsic jet speed. This is not only important for understanding the nature of specific objects, but also to directly test the paucity of these jets at high redshift when compared to theoretical predictions \citep{Haiman2004, volonteri2011}. Indeed, the VLBI properties of the first blazar detected at $z>6$, PSO J0309+27 \citep{belladitta2020} could reconcile this apparent observational--theoretical mismatch \citep{spingola2020}, demonstrating how important VLBI observations could be in directly testing galaxy formation and evolution models (see also \citealt{Mazzolari01.2026.SKA}).
Also, VLBI observations can often help determining if candidates are indeed blazars (e.g., \citealt{banados2025}) or jetted AGN seen at larger viewing angles (e.g., \citealt{Cao2017}). However, jet asymmetries and brightness enhancements can be also influenced by dense high-$z$ environments, requiring a multi-band approach for the complex classification of these objects (e.g., \citealt{Duncan2023, Roy2024, gloudemans2025_jet}).
Nevertheless, once their nature is confirmed, blazars play a crucial role in determining the total number of jetted AGN at any redshift (as $N_{\rm jetted} \simeq 2\Gamma^2$; \citealt{ghisellini2014, caccianiga2019} and references therein).

Among the physical properties measured in jetted AGN $z>4$, the estimate of $T_\mathrm{b}$ of high-$z$ blazars seems to be controversial (e.g., \citealt{Cao2017}).
The combination of the brightness temperature decreasing as a function of frequency (above the critical frequency of 9 GHz, as described in \citealt{Lee2014}) and the high variability of such objects can strongly affect the estimated values of $T_\mathrm{b}$, providing an explanation of the low values at high redshifts. The combination of these two effects may lead, ultimately, to an underestimate of the Doppler factor at rest-frame frequencies $\nu_{\rm rest-frame} > 10$ GHz, as the brightness temperature is intrinsically lower. In this context, VLBI observations of both SKA-Mid and SKA-Low with greatly improved sensitivity and angular resolution may be vital to determine the actual value of $T_\mathrm{b}$ and its frequency dependency at high fidelity.

\begin{figure}
    \centering
    \includegraphics[width=0.99\linewidth]{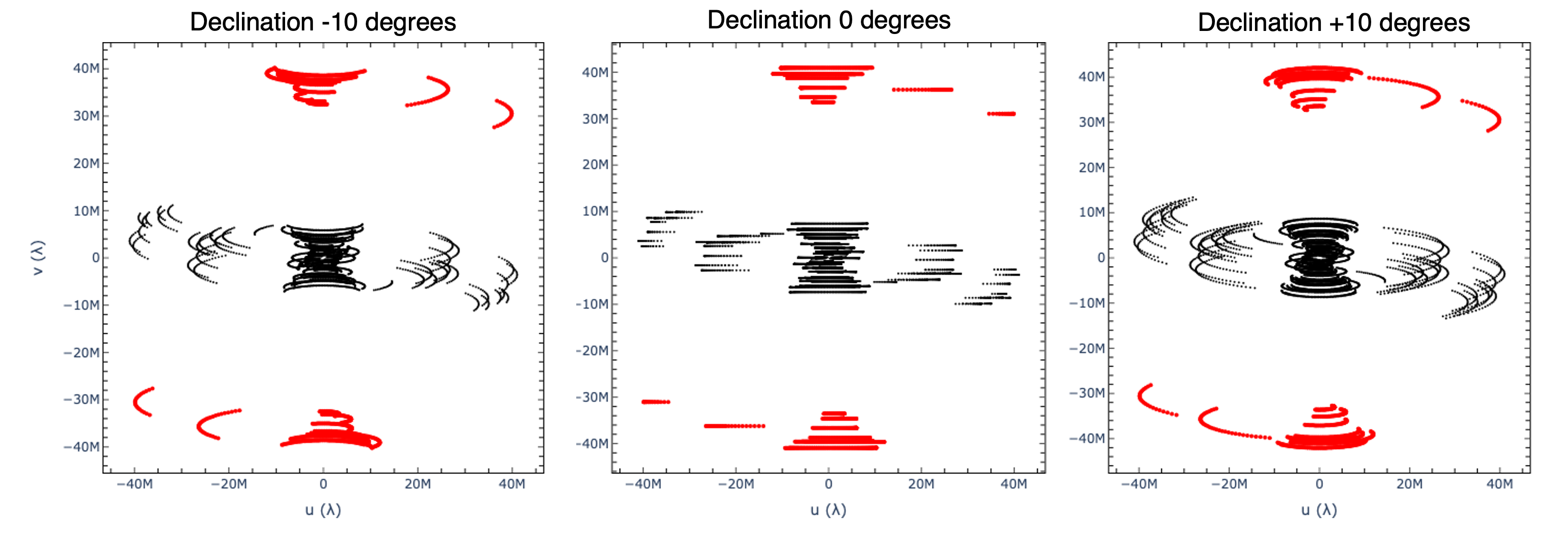}
    \caption{Simulated $uv-$coverage of a 12-hour observation at 1.4 GHz (Band 2) SKA-Mid AA4 with the European VLBI Network antennas (Effelsberg, Irbene, Jodrell Bank, Medicina, Noto, Sardinia, Onsala, Tianma, Torun, Urumqi and  Westerbork). In red we highlight the baselines to SKA-Mid AA4. The three panels show the $uv-$coverage for a source at a Declination of $-10$, 0 and $10$ degrees (from left to right, respectively). }
    \label{fig:uvcoverage}
\end{figure}

SKA-AA4 as a phased-array with the international baselines will already provide a sufficient sensitivity and angular resolution to detect and spatially resolve even faint jets at low surface brightness out to $z>6$. A simulation of the best possible $uv-$coverage (12-hour tracks) with the European VLBI Network antennas at different declinations is shown in Fig. \ref{fig:uvcoverage}. For these observations, the expected theoretical image sensitivity is of 1.3 $\mu$Jy beam$^{-1}$, which, paired with the well-filled $uv-$plane, will enable high image fidelity and high dynamic range images also of the faintest (hence, more common) jets (see Sec. \ref{sec:4} for quantitative estimates). Both SKA-Low and SKA-Mid will be important to fully characterize these high$-z$ sources in a complete multi-band view. 
A key observational subtlety is the $(1+z)$ frequency shift: an observing frequency of 1.4~GHz corresponds to $\simeq9.8$~GHz in the rest frame at $z=6$, biasing VLBI imaging toward more core-dominated emission. To recover the rest-frame $\sim1$--3 GHz regime where extended steep-spectrum jets and lobes are brighter, SKA-Low and SKA-Mid Band~1 become essential. 

In this context,  SKA-Low--VLBI can play a crucial role for determining the fraction and the spatially resolved properties of these objects, which are much needed to fully understand jetted AGN
(see also \citealt{Timmerman01.2026.SKA,Kobayashi01.2026.SKA}).

% SF
\section{Compact structures of AGN across the redshift space: evolution and cosmological manifestations}\label{sec:6}
Jetted AGN populate the entire observable Universe and, due to their enormous power, are observable from our cosmic neighborhood up to the highest redshifts ($z \gtrsim 6$). Consequently, they are ideal candidates as cosmological probes \citep[e.g.][]{2023AN....34430054R}. In comparison, Type Ia supernovae that are used as ``standardizable candles'' to infer cosmological model parameters are observed up to $z \lesssim 3$ only \citep{2024ApJ...971L..32P,2025A&A...701A..70V}. 
Since cosmological model predictions diverge most at high redshifts, having a cosmological probe such as jetted AGN can be of special value. Moreover, it is reasonable to assume that the physical processes of jet production operate in the same way across the widest range of redshifts, since the innermost pc-scale regions of AGN should be essentially free from the influence of the host galaxy and the intergalactic environment. 
This facilitates using compact jet structures as ``standardizable metric rods''. Indeed, earlier studies of their angular size--redshift \citep{kellerman1993,gurvits1994,gurvits1999, CaoDarkEnergy2017,Ghodsi2026} and apparent proper motion--redshift \citep{1994ApJ...430..467V} relations from high resolution VLBI observations showed that jetted AGN generally behave as expected from $\Lambda$CDM cosmologies. More recently, also variability from jetted AGN has been proposed as a method to measure the angular diameter distance up to high redshifts \citep{hodgson2020}.

The apparent angular extent of mas-scale jet structures observed with VLBI at the same frequency flattens out and even starts to increase with redshift at above $z \approx 1-2$ \citep{kellerman1993}. Whether the observed relation can be applied in practice for precise estimates of certain cosmological model parameters still remains to be seen. A crucial requirement is to increase the sample available for such studies, especially at the highest redshifts, in which SKA--VLBI can play an important role, thanks to its high sensitivity.

\begin{figure}[h]
    \centering
    \includegraphics[width=\columnwidth]{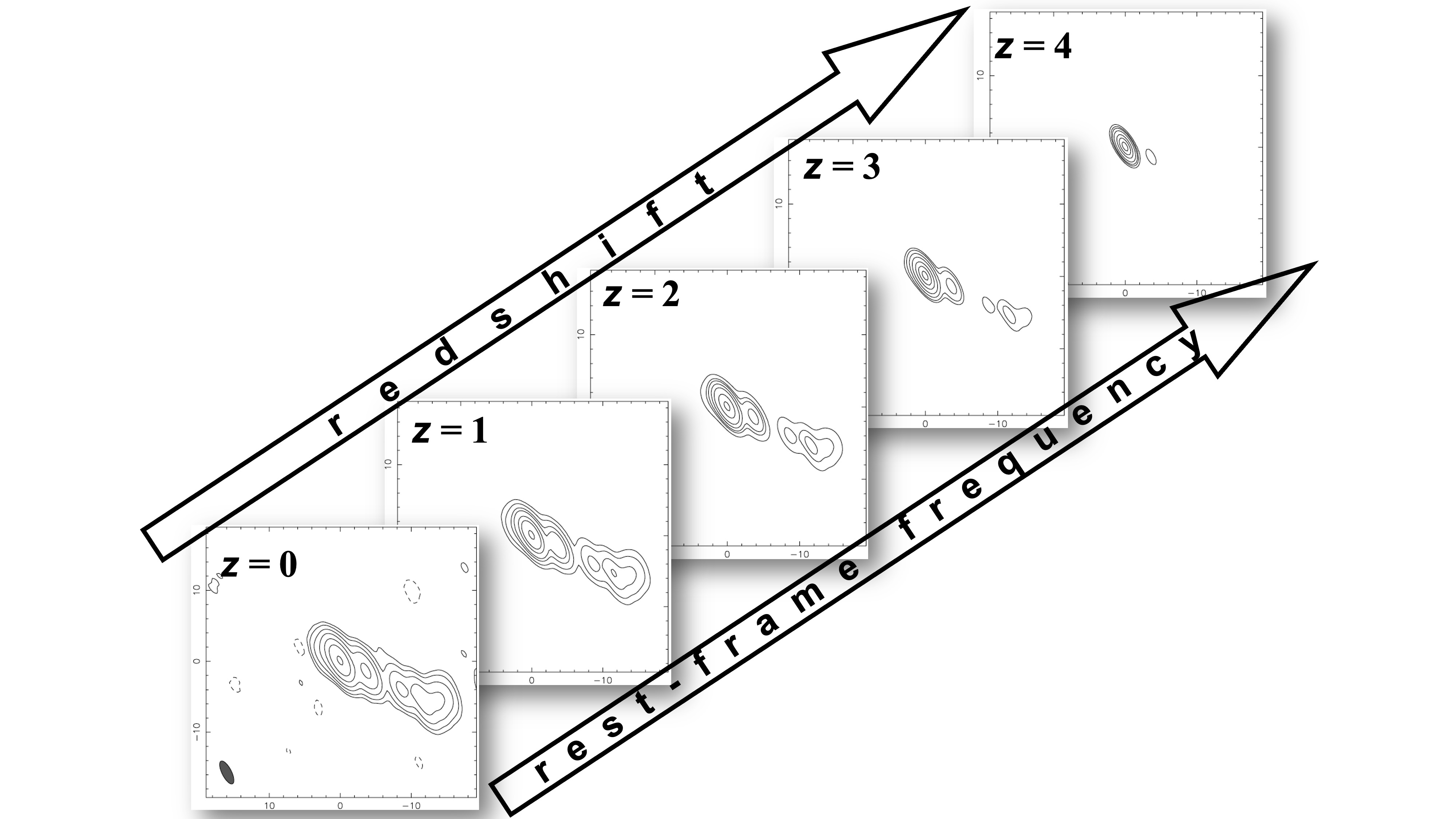}
    \caption{Simulated core--jet structure of a radio AGN at a given fixed arbitrary observing frequency (e.g. $1.6$~GHz), as moved to higher redshifts and consequently to $(1+z)$ times higher emitted frequencies. The steep-spectrum jet emission becomes gradually weaker compared to the flat-spectrum core at $(0, 0)$. The coordinate scales are in mas.}
    \label{fig:cjet-vs-z}
\end{figure}

The redshift dependence of the apparent jet proper motions observed via multi-epoch VLBI monitoring is also consistent with the concordance cosmological model: no extreme values \citep[i.e., exceeding about $0.2$~mas\,yr$^{-1}$;][]{2022ApJ...937...19Z} are measured at high redshifts, up to $z \approx 5.5$ \citep{2020NatCo..11..143A}. However, the sample of very distant AGN with VLBI jet proper motion measurements available to date is still very small, contains around $50$ objects, despite some recent efforts in analysing extensive archival data \citep{2025Univ...11...91G}. There are various observational limitations that so far prevented us from building up a sizeable AGN sample with reliable jet proper motion measurements. First of all, the total number of high-$z$ jetted AGN known to date is relatively small compared to lower-redshift samples. Moreover, in the expanding Universe, the emitted (rest-frame) frequency is related to the observed frequency as $\nu_\mathrm{em} = (1+z) \nu_\mathrm{obs}$. The high $\nu_\mathrm{em}$ at high $z$ means that the optically thin steep-spectrum jet emission gradually diminishes compared to the synchrotron self-absorbed core emission characterised by flat radio spectrum (Fig.~\ref{fig:cjet-vs-z}). In addition, extended radio emission at high redshifts is quenched due to inverse-Compton scattering on the CMB photons, making it more difficult to detect \citep{2017MNRAS.468..109W}.

There is one more way the Universe complicates our efforts to measure jet component proper motions: cosmological time dilation. This effect makes motions appear $(1 + z)$ times slower in the observer's frame compared to the rest frame of the source. The only remedy is to monitor high-redshift jetted AGN over extended periods of at least $5-10$~yr -- the earlier we begin, the better. In the other aspects, however, SKA--VLBI can be of immediate help. Low observing frequencies, coupled with superior sensitivity, increase the chance for detecting extended steep-spectrum mas-scale jets.  

An alternative way of ``mitigating'' the cosmological time dilation effect is extending VLBI baselines to extraterrestrial scales (i.e., space-VLBI arrays). Coupled with the superior sensitivity of the SKA-Mid, these systems would allow to exceed angular resolution limitations imposed by the Earth diameter. This is especially important for the discussed here science applications at lower frequencies, as illustrated by Fig.~\ref{fig:cjet-vs-z} 
\citep{gurvits2000,Kovalev01.2026.SKA}.

% YL
\section{Spatially resolved AGN jet feedback at high redshift}\label{sec:7}
%Yuanqi Liu \& her team

AGN feedback, whereby energy from the SMBH regulates its environment, is thought to play a key role in galaxy formation models \citep{2012ARA&A..50..455F}. In particular, relativistic jets launched from the AGN can carry kinetic energy into the host galaxy and beyond, potentially heating gas and suppressing star formation (negative feedback; e.g., \citealt{murthy2022}) or, in some cases, compressing gas to trigger star formation (positive feedback; e.g., \citealt{maiolino2017}). At early cosmic epochs ($z>6$), understanding the impact of jet-driven feedback is crucial, as it could influence the assembly of the first massive galaxies and the thermal state of the intergalactic medium. Yet, until recently this jet-induced feedback at high-$z$ has received relatively little attention (e.g., \citealt{best2014, migliocchetti2022,kondapally2023}). 
Studying AGN jet feedback at $z>6$ is vital for understanding early black hole–galaxy co-evolution and the state of the young Universe. 

The key limitation is not the physical importance of jets but the difficulty of identifying sufficiently bright, compact jetted AGN suitable for VLBI imaging and for matched multi-wavelength follow-up at $z\gtrsim6$.
Observationally, only a handful of AGN at $z>6$ have been detected and imaged on mas scales with VLBI. Moreover, only for a limited sample of these jetted AGN it was possible to access the molecular gas properties to test directly the presence of feedback from jets, which was found to be negligible \citep{khusanova2022}.
Extended lobes or secondary components tens to hundreds of mas from the core (tens to hundreds of parsecs at high-$z$) should become detectable, if present, and resolvable with the SKA--VLBI, as it can push below the $\mu$Jy-level surface brightness of such features. 
Far-reaching implications of resolving high-$z$ jets lie in the AGN--galaxy co-evolution field. With SKA--VLBI we will be able to observe how jets interact with their host galaxies at an epoch when galaxies themselves are in early stages of assembly. Are these jets punching out of their host and driving large-scale outflows, or are they stalling within dense proto-galactic gas? If SKA images reveal compact double-lobed structures at high-$z$, those would be young jets likely confined to the host galaxy’s central region – an indication that AGN feedback at that time might have been “buried” and affecting primarily the nucleus. Alternatively, if extended kpc-scale radio structures are found (analogous to mature radio galaxies), it would demonstrate that some AGN were capable of affecting the CGM and beyond even at $z>6$. 

Finally, SKA--VLBI observations can be combined with information from other wavelengths (e.g. ALMA maps of [C II] or CO lines tracing cold gas, or \textit{JWST} imaging of star formation) to assess the impact of jets on their hosts (as described in Sec. 8). Do jetted AGN at $z>6$ show quenched star formation or depleted gas in their host galaxies, as might be expected if jet feedback expels or heats the gas? Or conversely, is there evidence of jet-triggered star formation, such as alignment between radio jets and clumps of enhanced star formation (a positive feedback scenario)? At present, we lack of a statistically significant sample of $z\gtrsim6$ AGN hosts with confirmed jets and spatially resolved ISM diagnostics to answer these questions conclusively. Finding this sample and imaging it with SKA--VLBI will provide the final ingredient to assess how jets and early galaxies affected each other.

%CS & all
\section{Synergies with current and future telescopes}\label{sec:8}
%C. Spingola

A panchromatic investigation of AGN jets at $z>6$ is essential for understanding the early growth of SMBHs and their co-evolution with host galaxies during the EoR.
The ultra-sensitive, high-resolution radio imaging across a broad frequency range of SKA-VLBI will allow us to detect and morphologically study faint, high-redshift radio jets that are currently inaccessible (or extremely observationally time-consuming) with existing facilities. This will enable us to probe the more common
AGN population. Furthermore, the synergy of SKA-VLBI with multi-wavelength current and future facilities (described in detail below) will provide crucial information. This combined data will help us break degeneracies in the parameters related to the physical processes of AGN jets.

\subsection{Millimeter and sub-millimeter wavelengths} 

The 158\,$\mu$m fine-structure line of singly ionized carbon ([C\,\textsc{ii}]) is one of the most luminous and accessible tracers of the ISM in galaxies at high redshift. It arises primarily from photodissociation regions, but can also have contributions from the cold neutral and ionized gas phases. In the context of high-$z$ AGN and jet-hosting galaxies, [C\,\textsc{ii}] emission detected by ALMA
and/or the NOrthern Extended Millimeter Array (NOEMA), serves as a powerful diagnostic of both the ISM and jet--host interactions.

At $z > 6$, the [C\,\textsc{ii}] line is redshifted into ALMA Bands 5 to 7 (i.e., 275--211\,GHz for $z \sim 6$--8), where ALMA offers high sensitivity and spatial resolution. 
The [C\,\textsc{ii}] line has been robustly detected in numerous $z > 6$ quasar host galaxies using both ALMA and NOEMA observations
(e.g., \citealt{banados2015_noema, Venemans2017a, Venemans2017b, Decarli2018, walter2022}), with typical line luminosities in the range $L_{\mathrm{[C\,II]}} \sim 10^8$--$10^{10}\,L_\odot$.  In jetted AGN, [C\,\textsc{ii}] line mapping provides spatially resolved kinematics that can reveal the impact of mechanical feedback --particularly jet-induced outflows or turbulence-- on the host galaxy’s gas reservoir (e.g., \citealt{Neeleman2021}). Deviations from ordered rotation (e.g., broad, asymmetric, or multi-component profiles) can indicate jet- or AGN-driven turbulence and outflows. The radio synchrotron emission from SKA--VLBI may clearly reveal jet clearing or jet-triggered star formation (e.g., simulations by \citealt{Mukherjee2018}).

One of many benefits from the upcoming ALMA Wideband Sensitivity Upgrade (WSU) is the efficient spectral line observation of high$-z$ galaxies with serendipitously detected multiple lines in a single spectral tuning. The combination of the ALMA WSU and SKA-Mid will provide us with large spectral coverage, ensuring that we can observe several transitions and obtain the gas physical parameters via radiative transfer analysis or detect weak species via stacking techniques.

Finally, the future AtLAST \citep{Mroczkowski2025} would provide continuum sensitivity comparable to ALMA/WSU but with up to $10^5$ faster mapping speed once its focal plane is fully populated (see also \citealt{klassen2020}).

\subsection{Infrared and optical bands}

High angular resolution imaging and spectroscopy enabled by adaptive optics on large Southern Hemisphere facilities (e.g., VLT/ERIS, HAWK-I+GRAAL, and future ELT instruments such as MICADO and HARMONI) provide a crucial complement to SKA for studying high-$z$ AGN jets. At $z>6$, these systems probe rest-frame UV/optical emission (e.g., Ly$\alpha$) with $\lesssim 100$\,pc resolution, enabling detailed characterization of stellar structure, ionized gas, and jet–host interactions. In particular, ELT/MICADO’s diffraction-limited imaging ($\sim10$\,mas; \citealt{Davies2010}) will resolve compact hosts and merger signatures, while advanced AO instruments such as MAVIS will deliver near-diffraction-limited visible imaging and spatially resolved spectroscopy, allowing precise measurements of stellar mass, gas properties, star formation, and the dynamical state of quasar host galaxies.

\textit{JWST}’s NIRSpec and MIRI instruments provide rest-frame UV/optical spectra at unprecedented sensitivity, enabling robust redshifts, SMBH mass estimates via broad lines (e.g., H$_{\beta}$, H$_{\alpha}$), and detailed host galaxy stellar populations and star formation rates \citep{Bunker2023}. Follow-up observations with \textit{JWST} of SKA-identified AGN at $z>6$ could yield SMBH mass measurements, ionization diagnostics, host galaxy and environment properties. Integral field spectroscopy will spatially separate AGN and star-forming regions, informing models of radiative and mechanical feedback \citep{yang2023_jwst, yue2024, decarli2024}.

LSST’s deep, wide-area optical imaging and multi-band photometry will identify high-$z$ AGN candidates through dropout techniques, color selection and variability (e.g., \citealt{ivezic2019, belladitta2020}). Its time-domain survey enables optical variability studies on timescales from days to years, complementary to radio variability detected by SKA1-MID surveys, if obtained with multiple visits.

\textit{Euclid}’s deep, high resolution imaging and spectroscopic surveys over thousands of square degrees will provide complementary rest-frame optical/near-IR data, including host galaxy morphology, stellar mass estimates, and photometric redshifts for millions of galaxies. For AGN hosts, \textit{Euclid} spectroscopy can deliver emission line diagnostics (e.g., H$\alpha$, [O\,\textsc{iii}], \citealt{lusso2024_euclid}) and continuum shapes, essential for characterizing stellar populations and ionized gas properties. Moreover, \textit{Euclid}’s wide-area coverage aids the selection of exotic high-$z$ radio-loud AGN targets for SKA--VLBI follow-up, maximizing the scientific return from both facilities in probing the earliest epochs of jet activity.

\subsection{X-rays}

For about 25 years \textit{Chandra}’s sub-arcsecond angular resolution and high sensitivity has enabled the detection of inverse Compton X-ray emission from relativistic jets, which is boosted at high redshift by the $(1+z)^4$ increase in CMB energy density, allowing studies of jet energetics and particle acceleration (e.g., \citealt{schwartz2002}). Joint SKA/SKA--VLBI/\textit{Chandra} observations, potentially including deep ($\sim1$~Ms) pointings, can provide spatially resolved X-ray emission and spectral constraints on electron energy distributions and jet power, crucial for modeling AGN jet radiative processes and feedback.
Complementary, eROSITA is providing a nearly all-sky (only half of the sky will be public) catalog of bright X-ray sources in the $\sim 0.3-2$ keV band. While the sensitivity is not comparable to pointed observations with \textit{Chandra} and the angular resolution is poorer, it has been already successfully used to select and study bright blazars at Cosmic Dawn \citep{wolf2023,wolf2024}.

The large-area X-ray observatory "New Advanced Telescope for High-ENergy Astrophysics" (NewAthena) is  planned for launch in 2034 \citep{Cruise2025} and would allow us to constrain the overall SMBH accretion rate
density, reaching AGN around the knee of the luminosity function (where most black hole mass growth occurs). This would extend to the EoR for unobscured objects and characterize moderate to intermediate obscuration up to $z \sim 6- 7$, which may dominate accretion growth. 
This critical parameter space is challenging for facilities at other wavelengths to explore with sufficient statistics but the synergy with SKA will be important to detect a large number of high$-z$ AGN.

\subsection{Gamma-rays}  

Searches for $\gamma$-ray emission from high-$z$ blazars with the Fermi- \textsl{Large Area Telescope} (\textsl{LAT}) have relied on long-term data accumulation and the monitoring of flaring activity, but, at $\gamma$-ray energies, the K-correction strongly disfavors the detection of high-$z$ AGN. In addition, attenuation by the extragalactic background light (EBL) further suppresses the high-energy flux, leading to a severe negative K-correction that reduces the observed $\gamma$-ray brightness and hampers detections at $z\gtrsim 6$.

The highest-redshift source in the Fourth Fermi-\textsl{LAT} AGN catalog  remains GB~1508+5714 at $z=4.31$ \citep{Gokus2022}. Current strategies to identify high-$z$ $\gamma$-ray blazars typically begin with radio-selected samples (e.g., \citealt{kreter2020}; $z>2.5$). 
Selecting high-$z$ candidates at radio wavelengths as a first step is crucial because radio emission from AGN jets is largely unaffected by dust extinction, providing an efficient and unbiased way to identify distant sources. Moreover, radio selection (especially with VLBI) isolates the most powerful, beamed jets (i.e., blazar candidates) that are intrinsically more likely to produce detectable $\gamma$-ray emission despite strong K-correction and EBL attenuation, thereby maximizing the chances of successful $\gamma$-ray follow-up.  
This anticipates an important future synergy between SKA-VLBI and the Fermi-\textsl{LAT}.

\section{Acknowledgments}
CS acknowledges financial support by the Italian Ministry of University and Research (grant FIS2023$-$01611, CUP C53C25000300001) and by the INAF (through \textsl{Ricerca Fondamentale 2024}, Ob. Fu. 1.05.24.07.04).
MM acknowledges support from the Spanish Ministry of Science and Innovation through the project PID2024-159201NB-C22. This work was also partly supported by the Spanish program Unidad de Excelencia Mar\'ia de Maeztu CEX2020-001058-M, financed by MCIN/AEI/10.13039/501100011033, and by the MaX-CSIC Excellence Award MaX4-SOMMA-ICE.
TA acknowledges the Shanghai Oriental Talent Project.

\bibliographystyle{abbrvnat-maxbibnames4}
\bibliography{chapter} % if your bibtex file is called example.bib

\end{document}